Ferromagnetism and suppression of metallic clusters in Fe implanted ZnO – a phenomenon related to defects?

Shengqiang Zhou, K. Potzger, G. Talut, H. Reuther, K. Kuepper, J. Grenzer, Qingyu Xu, A. Muecklich M. Helm, and J. Fassbender

*Institute of Ion Beam Physics and Materials Research, Forschungszentrum Dresden-Rossendorf, P.O. Box 510119, 01314 Dresden, Germany*

E. Arenholz

*Advanced Light Source, Lawrence Berkeley National Laboratory, Berkeley, CA 94720 USA*

We investigated ZnO(0001) single crystals annealed in high vacuum with respect to their magnetic properties and cluster formation tendency after implant-doping with Fe. While metallic Fe cluster formation is suppressed, no evidence for the relevance of the Fe magnetic moment for the observed ferromagnetism was found. The latter along with the cluster suppression is discussed with respect to defects in the ZnO host matrix, since the crystalline quality of the substrates was lowered due to the preparation as observed by x-ray diffraction.

I. Introduction

Diluted magnetic semiconductors (DMS) are potential candidates for spintronics applications such as injection of spin polarized currents. Among others, transition metal (TM) doped ZnO is one of the candidates with a predicted Curie temperature above 300 K[1-2]. A comprehensive review on ferromagnetic ZnO is given in Ref. 3. Summarizing the previous experimental work, two interesting patterns emerge:

1. Very different magnetic order has been reported even for the same TM dopant including from paramagnetic and antiferromagnetic[4].

2. Ferromagnetic properties appear to be achievable in ZnO doped by almost every element. This includes such exotic dopants like, e.g. C[5].

The first observation suggests a high sensitivity of the magnetic properties on the preparation conditions. The second suggests that the development of ferromagnetic order is independent of the dopant. Intrinsic ferromagnetic properties, e.g. in defective C[6], are already well known. Consequently, the formation of ferromagnetic properties in ZnO without additional transition metal doping has been investigated very recently[7]. In Ref. 7, the authors applied energetic Ar ions of high fluences for defect built-up. The requirement of structural defects for ferromagnetism is supported in Ref. 8. On the other hand, possible unwanted contamination with iron[9] or other TM due to the production or handling process has to be critically discussed. A third possibility for the creation of ferromagnetic properties are secondary phases, i.e. metallic transition metals like Ni or Fe that are sometimes not detectable by standard structural analysis methods[10-12]. In a recent paper[13] we showed the possibility of suppression of the metallic secondary phases by means of annealing the ZnO single crystals in high vacuum prior to the implantation. Nevertheless, a very weak residual ferromagnetic signal could be observed. In this paper we discuss this suppression with respect to the defects created during preparation. Moreover, we show that the weak ferromagnetic signal observed is not related to the chemical presence of the implanted Fe but also appears in intentionally undoped samples. Thus it is either purely defect related or originates from contamination with transition metals that show abnormal annealing behaviour. Thus, we elucidate the origin of ferromagnetic order in low temperature $Fe^+$ implanted ZnO[10,13] observed earlier.

II. Experiment

As in ref. 13 we used commercial hydrothermal ZnO(0001) single crystals from the same sample charge. Prior further processing the samples have been annealed in high vacuum (base pressure < $1 \times 10^{-6}$ mbar) at 1073 K (for 30 min). The annealing process was performed on a sample holder made purely of Mo that was cleaned by several annealing cycles at maximum 1123 K. The lower annealing temperature as compared to ref. 13 was chosen to reduce the probability of contamination of the samples from the annealing process itself. Nevertheless, we performed superconducting quantum interference device magnetometry (SQUID) after test annealing of two ZnO virgin single crystals at 773 K and 1073 K, respectively. None of those, however, shows indications for ferromagnetic properties. We compared undoped and low temperature (maximum 253 K) $^{57}$Fe-implanted ZnO single crystals. An ion energy of 80 keV leads to a projected range of 38 nm and a straggling of 17 nm of the $Fe^+$ ions (TRIM). Post-annealing in high vacuum (base pressure < $10^{-6}$ mbar) was performed in isochronal temperature steps of 423 K, 573 K, 723 K and 773 K, respectively. The annealing time was 30 min. Magnetometry was performed after the 723 K and 773 K post-annealing step for each sample. A sample nomenclature is given in Table I. For each of the implanted Fe fluences, two samples have been prepared for later comparison. Magnetometry has been performed using SQUID (Quantum Design MPMS) with the magnetic field applied parallel to the sample surface. Electronic and structural analysis has been performed by means of conversion electron Mössbauer spectroscopy (CEMS), transmission electron microscopy (TEM) using a FEI Titan 80-300 st, X-ray magnetic circular dichroism (XMCD) at beamline 6.3.1 of the Advanced Light Source (ALS) in Berkeley, and high-resolution (HR) reciprocal space mapping X-ray diffraction (GE HXRD 3003).

III. Results

Immediately after implantation and post-annealing, magnetometry has been performed. The data presented in Fig. 1 indicates similar (but not completely equal) magnetic properties for undoped and Fe implanted samples, respectively. The zero field cooled (ZFC) and field cooled (FC) curves were obtained by initially cooling down the sample in zero field from 300 K to 5 K and subsequently warming it up in a field of 100 Oe while recording the magnetization (ZFC). After warming up the sample is cooled down again in the same field (FC). A bifurcation of both curves below a certain temperature $T_{Irr}$ is an indication to a magnetically induced irreversibility including ferromagnetic hysteresis. Without post-annealing, XX:ZnO, which also represents the precursor for the implanted samples, does not show a separation of both curves (Fig. 1a). As-implanted Fe(2.5%):ZnO and Fe(20%):ZnO show only marginal separation (Fig. 1b,d). By contrast, as-implanted Fe(10%):ZnO shows a pronounced irreversibility (Fig. 1c). $T_{Irr}$ is around 50 K. For XX:ZnO, ferromagnetic properties can be induced simply by means of vacuum post-annealing without any additional TM doping. Fig. 1a shows the ZFC/FC curves taken for XX:ZnO after the 723 K and 773 K annealing step, respectively. Moreover, a M-H hysteresis is observed after 773 K annealing (inset) in addition to the pronounced diamagnetic contribution. Subsequently, all of the implanted samples have been annealed using the same steps as for XX :ZnO. Creation or increase of ferromagnetic properties are observed in all of the implanted samples after the 773 K annealing step (Figs. 1 b - d). For samples XX:ZnO and Fe(2.5%):ZnO, test-like post-annealing at 823 K leads to a decrease of $T_{Irr}$ (Figs. 1 a,b). The irreversibility for all 773 K annealed samples gives rise to small hysteresis loops preserved up to room temperature (insets of Fig 1a - d). Note that the difference between the 5 K and the 300 K loop is less pronounced for XX:ZnO than for the Fe implanted sample. The latter suggests an additional effect of the implantation on the magnetic properties. The shape of the ZFC/FC curves as well as the large values of $T_{irr}$ is not typical for small spherical Fe nanoparticles. It could be explained by separated magnetic units of different kind with a broad anisotropy distribution. Consequently, no indication for the formation of crystalline Fe clusters has been found even for the sample with the largest Fe content, i.e. Fe(20%):ZnO, by means of XRD angular scans using a Siemens D5000 diffractometer (not shown) and TEM (Fig. 2).

The samples with the highest magnetic moment after the 773 K post-annealing step, i.e. Fe(10%):ZnO and XX:ZnO, were subjected to further detailed analysis. In order to directly clarify, whether Fe is electronically involved in the magnetic properties, we applied XMCD measurements to sample Fe(10%):ZnO within max. 2 weeks after the SQUID magnetometry. The corresponding XMCD spectra were recorded at 20 K (Fig. 3). No evidence for ferromagnetic order assigned to the implanted Fe could be detected.

On the other hand, the valence state of Fe is a mixture of $2^+$ and $3^+$ (comp. to ref.13). This was confirmed by bulk-sensitive CEMS recorded at room temperature. The $Fe^{2+}$:$Fe^{3+}$ relation strongly decreases upon 773 K annealing with respect to the as-implanted state, i.e. from 39.7%:60.3% to 26.2%:73.8% for Fe(10%):ZnO and even from 37.4%:62.6% down to 0%:100% for Fe(20%):ZnO. Thus, the annealing process oxidizes the implanted Fe. The CEMS spectra for all the Fe-implanted and post-annealed samples are shown Fig. 4 in comparison to as-implanted samples. The trend of oxidation with post-annealing is visible for all of the implanted fluences (Fig. 4 a-f).

An influence of the Fe-implantation on the ferromagnetic properties is indirectly given by the different shape of the ZFC/FC curves and magnetization temperature dependence of XX:ZnO and Fe(10%):ZnO, respectively (Fig. 1). Another difference between both samples can be found in the evolution of the magnetic properties with exposure to ambient conditions. I.e., after CEMS and XMCD, i.e. 2 weeks after preparation, we again applied SQUID magnetometry in order to identify possible degradation of the magnetic moment. Fig. 5 shows that almost no degradation occurred for Fe(10%):ZnO while the pronounced ferromagnetic properties for XX:ZnO disappeared. Thus, the presence of Fe or other implantation related effects appear to stabilize the defects responsible for the ferromagnetic properties.

At this moment there is no explanation why a pronounced bifurcation is reached for a Fe concentration of 10% and not for the others. Due to the observed high fragility of the preparation process, a more direct control of the defect builtup would be necessary. In our case, at least the lowering of the crystalline quality by the vacuum pre-annealing and further preparation can be measured by means of a reciprocal space mapping (RSM) measurement at the (0004) ZnO reflection. The latter measurements have been performed after XMCD and CEM spectroscopy. The diffuse scattering of the processed samples (Fig. 6a,b) is much stronger than that of a comparable as-purchased virgin sample. Such behaviour usually indicates the development of strain and grain boundaries. A pronounced strain can be detected especially for Fe(10%):ZnO. Note that the FWHM in Fig. 6a is strongly increased and splitted as compared to Fig. 6b or 6c suggesting mosaicity increase. The enhancement of defects due to the implantation is supported by Fig. 2, i.e .TEM micrographs of the near surface region of Fe(20%):ZnO. In the implanted region down to 60 nm below the surface a large amount of planar defects is visible.

Besides the defects, the number of mobile charge carriers as potential carriers of the magnetic moments are expected to be altered due to the preparation. Hall measurements at 290 K, again for Fe(10%):ZnO, indeed show a mobility of 77.6 $cm^2V^{-1}s^{-1}$ and a charge carrier concentration of $8.74 \times 10^{18}$ $cm^{-3}$. The numbers are given with respect to a film thickness of 34 nm deduced from the ion straggling. Note the peculiarity that the Zn face of the post-annealed XX:ZnO sample is nearly insulating. Thus, the pronounced electrical conductivity in Fe(10%):ZnO is not the prior origin of the ferromagnetic coupling but might alter it.

IV. Discussion

For interpretation of the collected data, two questions have to be focused on. First, the origin of the weak ferromagnetic properties observed for all of the samples including

XX:ZnO without any implantation of Fe has to be discussed. Second, the mechanism of the suppression of the metallic secondary phases has to be explained. For the first we want to clearly point out that the annealing behaviour of the magnetization properties is not monotonic but reaches a maximum after annealing at 773 K and clearly drops after annealing at 823 K. Recent papers dealing with TM doped ZnO based diluted magnetic semiconductors, often explain the exchange mechanism of magnetic coupling by Zn interstitial induced n-type charge carriers[17]. Zn interstitials can be created by both ion implantation and annealing at moderate temperatures[18]. They are, however, not stable at high temperatures[19-20]. This behaviour would qualitatively agree with our magnetometry measurements on Fe implanted ZnO but could not explain the absence of a dichroic signal nor the similar behaviour between Fe:ZnO and XX:ZnO. The different stability upon exposure to air of Fe:ZnO and XX:ZnO might – on the other hand - be related to additional defects created from the Fe ion implantation. A scenario explaining the observed magnetic properties involving contamination from the production process is, however, difficult to establish but not impossible. In case of contamination with Fe ions, applying energy due to implantation or mild annealing cannot reduce these ions to metallic Fe if they are embedded (Fig. 4). The situation in the surface layers or for other TM might, however, be different. Decreasing $T_{irr}$ after annealing at higher temperatures again might be due to the oxidization of the reduced TM at the surface. The possibility of reduction-reoxidation of embedded Fe clusters in an oxide, i.e. yttria stabilized zirconia, has been proven earlier[21].

Turning to the second question, the suppression of metallic secondary phases by means of vacuum pre-annealing has to be connected with defects created due to such treatment (Fig. 6). The gettering of Fe as an unwanted contamination of, e.g. Si, by defects was studied already earlier. It became necessary due to the negative effect of interstitial Fe or Fe-Si complexes on the performance of computing devices or solar cells. From these studies it is known that unwanted Fe can form a huge variety of compounds with Si and oxygen in silicon technology. ZnO, on the other hand, consists of two elements with a large decomposition tendency upon annealing in reducing atmosphere. The crucial point, however, is the different decomposition behaviour upon annealing of ZnO material with different initial crystalline quality[23-24]. In ref. 23 it was shown that ion implanted ZnO shows a larger decomposition tendency as compared to as grown films upon annealing at the same temperature. In our case the crystalline quality was initially reduced due to vacuum pre-annealing (Fig. 6). Therefore a higher decomposition tendency can be expected as compared to as-purchased crystals. From that one can deduce that reduction of ZnO due to presence of Fe occurs at lower annealing temperatures. This leads to the formation of non-magnetic Fe-Zn-O complexes rather than metallic Fe clusters. Such mechanism becomes likely after comparing the CEM spectra from Fig. 4d and f to the ones from Fig. 3 c and d in Ref. 11. Obviously, the spectra are similar indicating similar oxidation behaviour of Fe in defective and non-defective ZnO but at different annealing temperatures. The lower annealing temperatures applied to Fe:ZnO, however, do not allow the formation of crystalline and thus ferromagnetic $ZnFe_2O_4$ as in ref. 11. Considering the pre-annealing induced defects, e.g. grain boundaries, as gettering centers for Fe, the presence of a large amount of such centers also suggests a large number of Fe ions immobilized in those complexes. This would explain the absence of metallic Fe clusters also in the highly Fe doped Fe(20%):ZnO. An incorporation of a part of the implanted Fe ions into the defective ZnO crystal, however, can not be excluded. One indication is the formation of a large fraction of $Fe^{2+}$ as compared to non pre-annealed crystals[13].

V. Conclusions

In conclusion we found that ferromagnetic order can be induced in ZnO(0001) single crystals by means of Fe ion implantation as well as vacuum annealing at mild temperatures without transition metal doping. Thus, speculations on a possible magnetic coupling of localized d-moments of the implanted Fe via charge carriers created by point defects like Zn interstitials (e.g. in Ref. 10, 13), i.e. the formation of a ferromagnetic DMS, are shown to be misleading. The weak ferromagnetic properties are likely defect induced but also might originate from contamination from the production process. The origin of the suppression of metallic secondary phases is explained from the formation of non-magnetic Fe-Zn-O complexes at defects.

Acknowledgement:

Partial financial funding (Q.X.) from the Bundesministerium für Bildung und Forschung (FKZ03N8708).

The Advanced Light Source is supported by the Director, Office of Science, Office of Basic Energy Sciences, of the U.S. Department of Energy under Contract No. DE-AC02-05CH11231.

Table I: Sample nomenclature

| Sample identifier | Fe fluence (x10$^{16}$ cm$^{-2}$) | Maximum Fe concentration |
|---|---|---|
| XX :ZnO | 0 | 0 at.% |
| Fe(2.5%):ZnO | 1 | 2.5 at.% |
| Fe(10%):ZnO | 4 | 10 at.% |
| Fe(20%):ZnO | 8 | 20 at.% |

Figure captions:

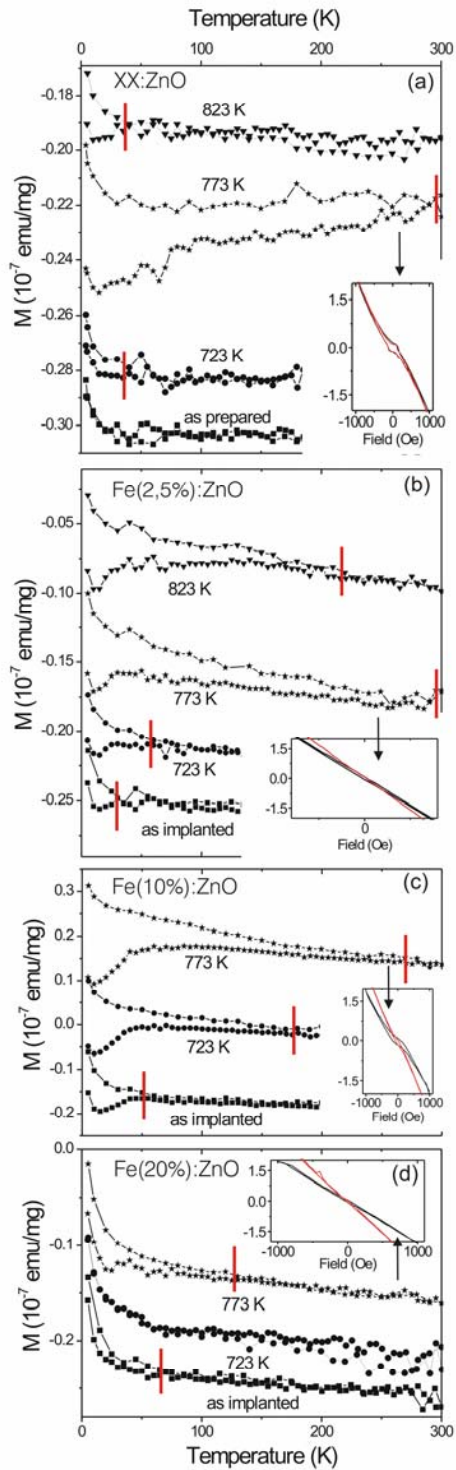

Fig. 1. ZFC/FC curves and M vs. H dependence (insets) for the various crystals after selected preparation steps. The samples are labelled according to Table I and the annealing temperatures are indicated. Vertical lines in the ZFC/FC curves mark the bifurcation points $T_{Irr}$. The spectra – except the as-prepared/as-implanted - are shifted in y-direction for better visibility. Black and red curves (insets) represent 5 K and 300 K measurement temperature, respectively. An unknown, likely device related y-offset has been subtracted for the insets.

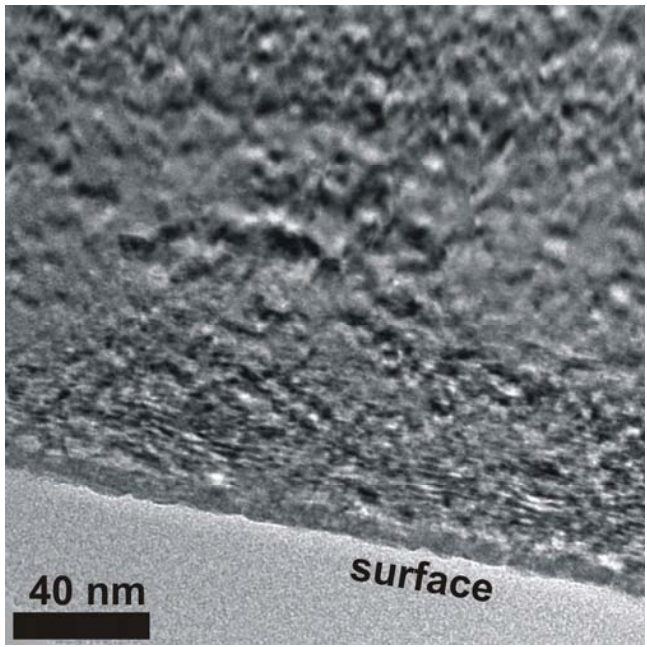

Fig.2 TEM micrograph of the near surface region of sample Fe(10%):ZnO after implantation and 773 K post annealing. No Fe secondary phases are visible but a large amount of planar defects in the implanted region (down to ~ 60 nm). Similar defects have been observed earlier[14].

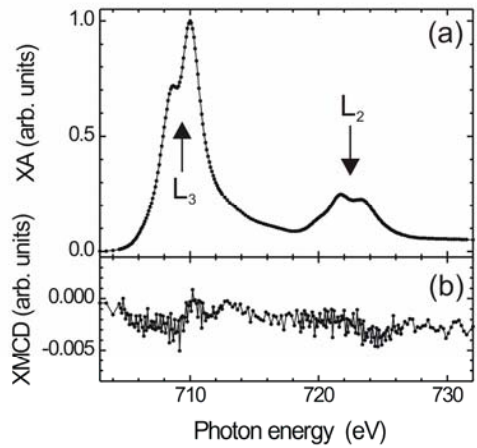

Fig. 3. (a) XA and (b) XMCD spectra of Fe(10%):ZnO. The magnetic field was switched between +2 kOe and –2 kOe for every data point at fixed polarization of the X-rays. The X-ray angle of incindence angle was 60° to the sample surface normal. The slight differences in (b) do likely not originate from an XMCD effect but from non-symmetric effects described by E. Goering for TEY-measurements in applied magnetic fields[15].

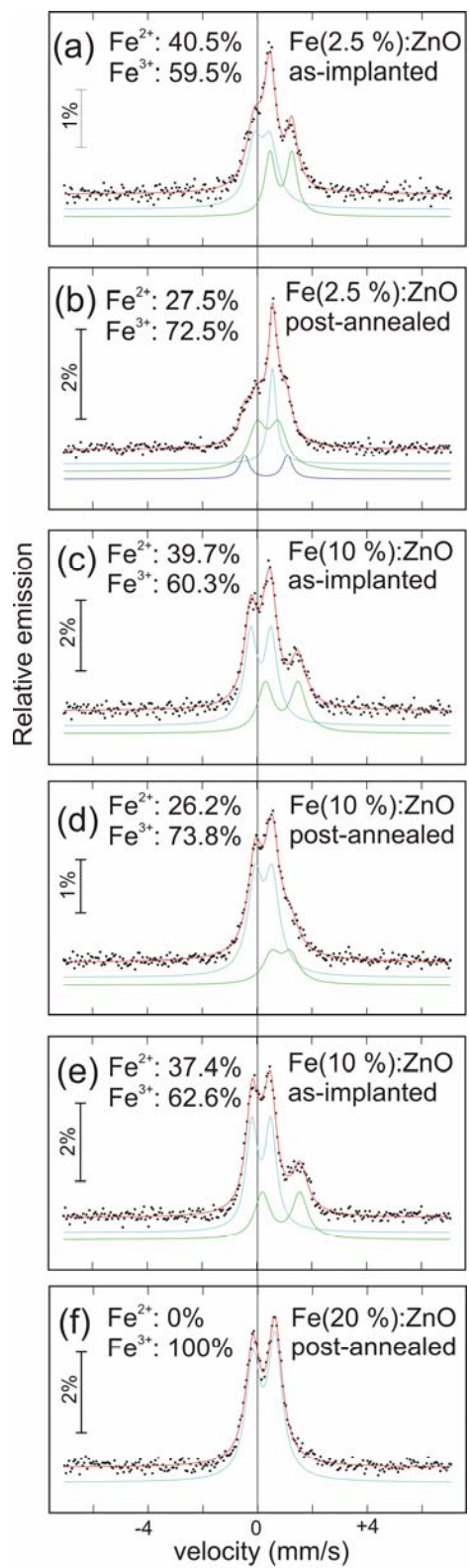

Fig. 4. CEM spectroscopy for all Fe implanted samples after implantation and post-annealing, respectively. All the samples have been implanted and measured

approximately at the same time. The spectra have been fitted using the NORMOS program[16]. The relation between the amounts of Fe ions in 2+ and 3+ oxidation states has been indicated.

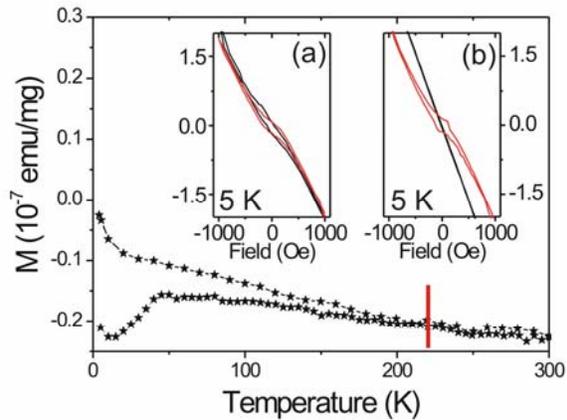

Fig. 5. ZFC/FC curve of sample Fe(20%):ZnO 2 weeks after preparation. Inset (a) shows the corresponding hysteresis (black) in comparison to the one taken direct after preparation (red). Inset (b) shows the corresponding comparison for XX:ZnO.

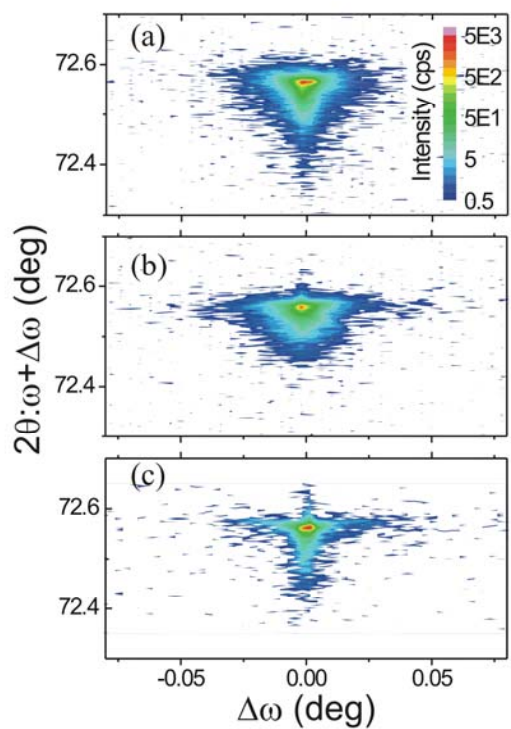

Fig. 6. HR-XRD at the ZnO(0004) reflection of (a) Fe(10%):ZnO, (b) XX:ZnO, and (c) a virgin sample from the same charge. (a) and (b) show a similar diffuse background in the reciprocal space that is much more pronounced than for a virgin sample (c).